# Evolution of correlated electronic states of La$_2$NiO$_4$ under hydrostatic pressure

Shu-Hong Tang[a,b,#], Han-Yu Wang[a,b,#], Da-Yong Liu[c], Feng Lu[d], Wei-Hua Wang[d], H.-Q. Lin[e] and Liang-Jian Zou[a,b,*]

a. Key Laboratory of Materials Physics, Institute of Solid State Physics, HFIPS, Chinese Academy of Sciences, Hefei 230031, China
b. Science Island Branch of Graduate School, University of Science and Technology of China, Hefei 230026, China
c. Department of Physics, School of Sciences, Nantong University, Nantong 226019, China
d. Department of Electronic Science and Engineering, and Tianjin Key Laboratory of Efficient Utilization of Solar Energy, Nankai University, Tianjin 300350, China
e. Physics College, Zhejiang University, Hangzhou 230031, China

**Abstract**

We elucidate the electronic structure and quantum many-body instabilities of the monolayer nickelate La$_2$NiO$_4$ under hydrostatic pressure using a combination of density functional theory, dynamical mean-field theory (DFT+DMFT), and random phase approximation (RPA). Our DFT+DMFT calculations reveal non-Fermi-liquid behavior and coherence loss near the Fermi level at low pressures, driven by strong electron correlations within the Ni-$e_g$ orbital manifold, which is analogous to the low-energy electronic properties observed in La$_3$Ni$_2$O$_7$. However, multi-orbital spin susceptibility analysis demonstrates an exceptionally suppressed critical Stoner parameter $U_c$ (~0.4–0.7 eV), indicating robust magnetic order that dominates the ground state and precludes superconductivity in the pristine system. Below $U_c$, superconducting instabilities exhibit a pressure-driven symmetry transition: $d_{x^2-y^2}$-wave pairing prevails at ambient pressure, while a transition to $s + g$-wave symmetry occurs above 75 GPa. This transition is attributed to pressure-induced self-doping effect. The high-angular-momentum $g$-wave component incurs significant energetic penalties, rendering high-Tc superconductivity unlikely. We conclude that the absence of superconductivity in La$_2$NiO$_4$ arises from its robust intrinsic magnetism and the unfavorable pairing symmetry under pressure, suggesting that alternative routes—such as chemical doping or epitaxial strain—are necessary to suppress magnetism and unlock superconducting states.

\# These authors contribute equally.

\* Corresponding author: zou@theory.issp.ac.cn

**Introduction** Recent discovery of unconventional superconductivity in pressurized bilayer La$_3$Ni$_2$O$_7$ (>14 GPa, Tc ≈ 80 K) [1-9] extends high-temperature superconductivity platforms beyond cuprates and iron-based systems to layered nickelates. While the search for superconductivity in related nickelates has intensified, observations reveal stark contrasts: trilayer La$_4$Ni$_3$O$_{10}$ exhibits diminished Tc [10-17], and monolayer La$_2$NiO$_4$ shows no superconductivity from ambient pressure to 50 GPa [18,19]. Currently, the markedly divergent behavior exhibited by these three materials under pressure remains poorly understood. Given the crucial role of electron correlations in the superconductivity of La$_3$Ni$_2$O$_7$ [20-28], investigating the differences in electronic correlations within La$_2$NiO$_4$—particularly their evolution under pressure and their implications for superconducting potential—is essential for understanding the emergence of high-temperature superconductivity in this new class of nickel-based superconductors.

Here, we study the electronic properties evolution of La$_2$NiO$_4$ under pressure by adopting a comprehensive density functional theory combined dynamical mean-field theory (DFT+DMFT) [29-33]. We derive precise tight-binding Hamiltonians from DFT bands and employ the random-phase-approximation (RPA) formalisms [34] to investigate spin-fluctuation-mediated pairing instabilities. Our results demonstrate that, despite lacking apical-oxygen-mediated coupling, La$_2$NiO$_4$ exhibits low-energy electronic structure topology strikingly similar to La$_3$Ni$_2$O$_7$ [35-38] and manifesting flat-band features and non-Fermi-liquid behavior near Fermi level. RPA calculations reveal an exceptionally suppressed critical Stoner instability interaction strength $U_c$, indicative of robust magnetism. Moreover, pressure—while potentially inhibiting magnetism—shifts the dominant pairing symmetry from energetically favorable $d_{x^2-y^2}$-wave to $s + g$-wave, where high-angular-momentum $g$-wave components incur significant energetic penalties that likely preclude high-Tc superconductivity.

**Crystal Structures** In experiments, La$_2$NiO$_4$ adopts the Ruddlesden-Popper structure A$_{n+1}$TM$_n$O$_{3n+1}$ with space group I4/mmm, No. 139 at room temperature and ambient pressure, as confirmed by high-resolution X-ray and neutron diffraction studies [39,40]. This layered perovskite structure consists of alternating LaO rock-salt layers and LaNiO$_3$ perovskite-like slabs stacked along the crystallographic c-axis, as illustrated in Fig 1a. The primitive unit cell contains two formula units, with lattice parameters a ≈ 3.82 Å and c ≈ 12.92 Å at ambient pressure [39,40]. A key structural feature is the presence of apical oxygen displacements along the *c*-axis. This distortion arises from the mismatch between the ideal Ni-O bond length and the lattice constraints, leading to bond length disproportionation (e.g., in-plane Ni-O(1) ≈ 1.91 Å vs. apical Ni-O(2) ≈ 2.34 Å. This raises the energy of $d_{x^2-y^2}$ relative to $d_{z^2}$ by coupling to oxygen displacement modes, resulting in robust orbital polarization confirmed by the XAS and the DFT+U method [41,42].

In contrast to the structurally related superconductor La$_3$Ni$_2$O$_7$, the La$_2$NiO$_4$ crystal structure, lacking apical-oxygen-mediated electron hopping, exhibits a more complex orbital composition in proximity to the Fermi level.

This complexity arises from competing effects: level splitting favors the adoption of a low-spin state, whereas orbital correlations conversely promote a high-spin configuration. Ultimately, this competitive interplay results in an electronic ground state exhibiting pronounced sensitivity to subtle orbital correlations. Consequently, elucidating the precise nature of this ground state necessitates the application of higher-accuracy electronic structure calculation methodologies.

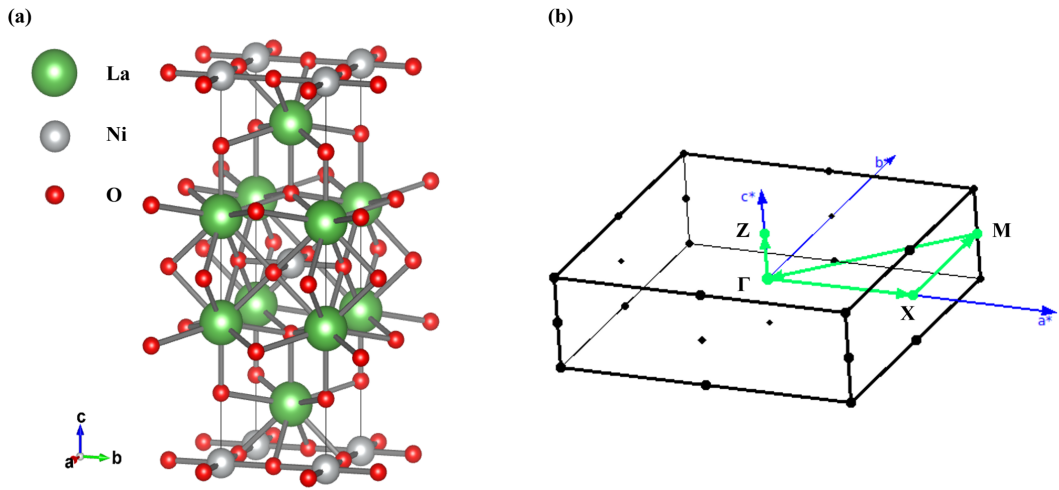

Fig. 1 (a) Crystal structure of $La_2NiO_4$, space group I4/mmm. (b) Conventional Brillouin zone of $La_2NiO_4$, with high-symmetry paths and points labeled for calculations.

**Electronic Structures under Various Pressures** To investigate the pressure-induced evolution of the electronic structures in $La_2NiO_4$, we obtained the k-resolved spectra at various pressures using the first-principles calculations. Fig 2 presents the DFT band structures and the Fermi surface of $La_2NiO_4$ at various pressures, serving as the starting point for our analysis. The fat-band analysis across all pressure regimes reveals that the low-energy physics near the Fermi surface is dominated by Ni-$3d$ orbital contributions, manifested as two hybridized bands formed by the two eg orbitals. This two-band feature corresponds to the single Ni site per unit cell in $La_2NiO_4$. As seen in Fig. 2(a-d), the two bands generate an electron pocket predominantly composed of $d_{x^2-y^2}$ orbital character near the Γ point, and a hole pocket with dominant $d_{z^2}$ orbital contribution around the X point in the Brillouin zone. The band structures near the Fermi level and the resulting Fermi surface show close resemblance to both the ARPES experimental observations and DFT computational results reported for $La_3Ni_2O_7$ [35, 44, 45]. Notably, both Fermi surfaces in $La_2NiO_4$ exhibit a finite hybridization between Ni-$d_{z^2}$ and $d_{x^2-y^2}$ orbital components, reflecting interorbital mixing effects, which is a characteristic consistent with observations in the high-temperature superconductor $Ba_2CuO_4$ under high pressure [46, 47]. The remaining Ni-$3d$ $t_{2g}$ orbitals exhibit negligible hybridization with the bands near the Fermi level. Their spectral weight is primarily distributed approximately 2 eV below the Fermi energy, thus making minimal contributions to the low-energy physics, as shown in Fig. 2.

Once applying hydrostatic pressure, the pressure-dependent band structures are shown in Fig. 2(a-c) respectively. One finds that both bands near the Fermi level exhibit significant broadening with increasing pressure, corresponding to the enhancement of intralayer intraorbital hopping parameters under pressure. Concurrently, the hybridization between $d_{x^2-y^2}$ and $d_{z^2}$ orbitals is further strengthened, as the interorbital coupling predominantly originates from inter-site hopping between adjacent orbitals, which is likewise amplified under pressure. Simultaneously, we observe the emergence of a new electron pocket around the Z point contributed by La-$f$ states at pressures exceeding 25 GPa, which exhibits progressive enlargement with increasing pressure. This development induces a redistribution of electron occupation within the Ni-3$d$ $e_g$ orbitals as seen in Fig. 2(d). Although $f$-electrons are typically localized and rarely participate in charge transport, the formation of this additional electron pocket effectively acts as a self-doping mechanism that modifies the occupation configuration of the 3$d$ $e_g$ orbitals. This modification, despite the inherent localization of $f$-electrons, significantly alters the correlated electronic structure of the 3$d$ states, an effect whose implications for electron correlation will be systematically examined in the following discussion of the electronic correlation effects.

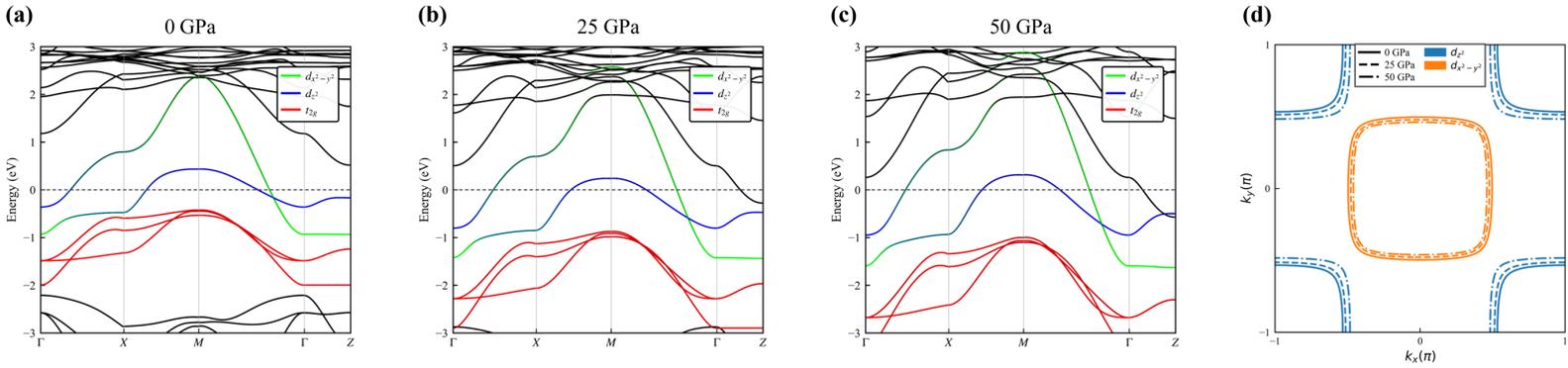

Fig. 2 DFT band structures of La$_2$NiO$_4$ at the pressure 0(a), 25(b), and 50 GPa(c) with orbital-projected contributions color-coded as $d_{z^2}$ (blue), $d_{x^2-y^2}$ (green), and $d_{t_{2g}}$ (red); (d) Momentum-resolved Fermi surface at $k_z = 0$ for 0, 25 and 50 GPa, colored by dominant orbital character matching the band structure scheme.

Based on DFT-derived band structures, we construct a two-orbital tight-binding model targeting the Ni $e_g$ orbitals:

$$H_{TB} = \sum_{ij,ll'\sigma} t_{ij,ll'} c^+_{il\sigma} c_{jl'\sigma} \quad (1)$$

Here, the operator $c^+_{il\sigma}$ creates an electron with spin $\sigma$ on orbital $l$ at lattice site $i$, where $l$ designates the $d_{x^2-y^2}$ and $d_{z^2}$ orbitals, respectively. The values of the hopping parameters $t_{ij,ll'}$ under different pressures were systematically computed using the Wannier90 software package interfaced with DFT calculations [43]. The resulting parameter values are summarized in Table 1 below.

Table 1. Tight binding hopping parameters $t_{ij,ll'}$ under varying pressures. Superscripts 0,1,2,3 denote on-site, first-, second-, and third-neighbor terms, respectively. Subscripts $x$ and $z$ correspond to the $d_{x^2-y^2}$ and $d_{z^2}$ orbitals, while $xz$ indicates interorbital

hopping. Due to orbital symmetry, interorbital hopping amplitudes along x- and y-directions exhibit opposite signs.

|         | $t_x^0$ | $t_x^1$ | $t_x^2$ | $t_x^3$ | $t_z^0$ | $t_z^1$ | $t_{xz}^{1,x}$ | $t_{xz}^{1,y}$ |
|---------|---------|---------|---------|---------|---------|---------|----------------|----------------|
| 0 GPa   | 0.362   | -0.398  | 0.071   | -0.020  | -0.091  | -0.072  | -0.163         | 0.163          |
| 25 GPa  | 0.484   | -0.477  | 0.076   | -0.032  | -0.111  | -0.090  | -0.200         | 0.200          |
| 50 GPa  | 0.576   | -0.533  | 0.078   | -0.042  | -0.108  | -0.106  | -0.229         | 0.229          |
| 75 GPa  | 0.663   | -0.579  | 0.079   | -0.058  | -0.0949 | -0.121  | -0.253         | 0.253          |
| 100 GPa | 0.750   | -0.621  | 0.080   | -0.063  | -0.080  | -0.135  | -0.275         | 0.275          |

Given that the bands near the Fermi level are exclusively composed of $3d$ orbitals, the Coulomb correlations among Ni-$3d$ electrons play a crucial role in determining the electronic structure. The dominant $3d$-orbital character near Fermi level necessitates explicit treatment of electron correlations, as captured by our DFT+DMFT framework incorporating Hund's coupling $J_H$ = 0.5 eV and Hubbard $U$ = 5.0 eV for Ni-$3d$ states. Our charge self-consistent DFT+DMFT calculations reveal substantial correlation-driven modifications in the low-energy electronic structure near the Fermi level under pressures, as comprehensively shown in Fig. 3(a-c). The low-energy physics is exclusively governed by $3d\ e_g$ orbital bands, while $3d\ t_{2g}$ bands reside deep below the Fermi level and contribute negligibly. Near the Fermi surface, band dispersions exhibit extreme flattening: the $d_{z^2}$ orbital forms a quasi-flat band, and the $d_{x^2-y^2}$ bandwidth is significantly narrowed by correlations as seen in Fig. 3(d) and (e). Crucially, a La-derived self-doping band crosses Fermi level near the Z-point at ambient pressure—different form DFT predictions—and penetrates deeper under compression, thereby modulating $3d\ e_g$ orbital occupancy. This drives the $d_{x^2-y^2}$ orbital toward proximate half-filling with $n(d_{x^2-y^2}) \approx 1.03$, thus amplifying the correlation effects of $La_2NiO_4$.

At ambient pressure, while dispersions retain coherent renormalization away from Fermi level, their integrity dissolves in the immediate vicinity of the Fermi level for both $d_{z^2}$ and $d_{x^2-y^2}$ orbitals, which progressively reverses under pressure of 20 GPa. This phenomenon demonstrates that correlations drive deepened $3d\ e_g$ orbital hybridization at the Fermi surface—a phenomenon similarly observed in $La_3Ni_2O_7$ [48]. The effect diminishes with pressure-induced reduction in $U/W$, the correlation-to-bandwidth ratio. Such behavior contradicts conventional Fermi liquid paradigms, indicating possible non-Fermi liquid formation under correlations at low pressures. Defining stable quasiparticles becomes untenable in bands where coherence remains irreconcilable.

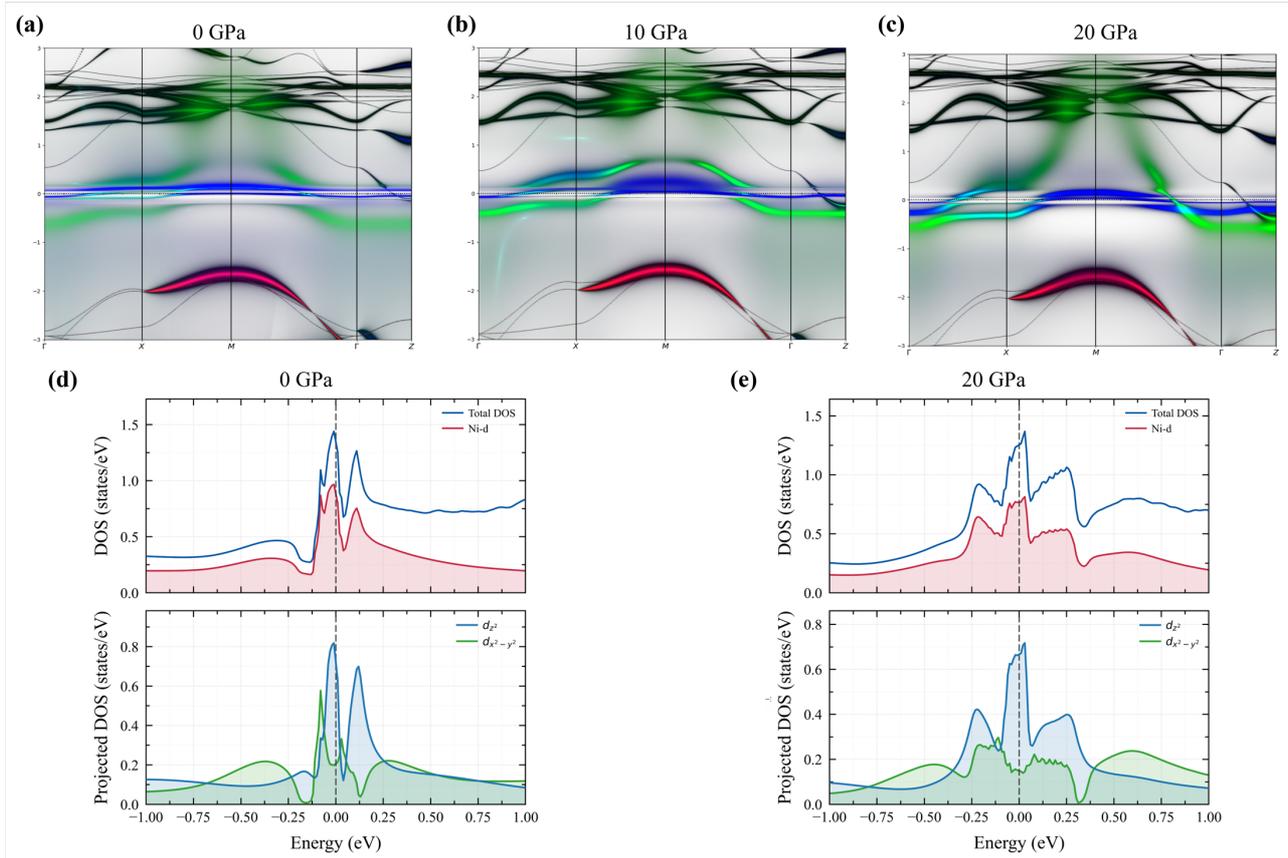

Fig. 3 (a-c) DFT+DMFT band structures computed with $U$=5eV and $J$=0.5eV under pressures of 0, 10, and 20 GPa, respectively. Color-coded by orbital-projected contributions: $d_{z^2}$ (blue), $d_{x^2-y^2}$ (green), and $d_{t_{2g}}$ (red); (d,e) Orbital-resolved density of states (DOS) at 0 GPa and 20 GPa, respectively, calculated with identical interaction parameters.

**Pressure-Dependent Magnetic Instabilities in La₂NiO₄** We also investigate the magnetic instability of La₂NiO₄ under different pressures from a weak-coupling perspective. Based on a two-band tight-binding model derived from DFT-calculated electronic structures, we construct an interacting Hamiltonian that incorporates electron-electron interactions:

$$H = H_{TB} + H_{int} \qquad (2)$$

The on-site interaction Hamiltonian $H_{int}$ adopts the Slater-Kanamori form. It encompasses intra-orbital and inter-orbital Coulomb interactions, spin-flip terms, and pair-hopping terms, parameterized by $U$, $U'$, $J_H$, $J_P$. To investigate magnetic instabilities and the potential for superconductivity within the system, we performed standard multi-orbital random phase approximation (RPA) calculations, which is described in supplemental material, on the interacting Hamiltonian in Eq. 2. Our approach involved first computing the multi-orbital RPA susceptibility under varying hydrostatic pressures to characterize the magnetic instability landscape of La₂NiO₄. The critical Hubbard interaction parameter $U_c$, signifying the onset of Stoner magnetic instability within this system, was subsequently determined by the value of $U$ at which $\chi_{RPA}(q, \omega = 0)$ diverges.

Table 2. The critical Stoner instability interaction strength $U_c$ under varying pressures, which is evaluated with $J_H = J_P = 0$.

|  | 0 GPa | 25 GPa | 50 GPa | 75 GPa | 100 GPa |
|---|---|---|---|---|---|
| $U_c$ / eV | 0.421 | 0.493 | 0.590 | 0.615 | 0.724 |

As summarized in Table 2, the calculated critical Stoner instability interaction strength $U_c$ for the La$_2$NiO$_4$ system at temperatures representative of our simulation regime (T = 50 K) ranges between 0.4 eV and 0.7 eV across several applied pressures. This observed $U_c$ range is notably small—approximately an order of magnitude lower—when compared to the characteristic tight-binding bandwidth of the system, which is approximately 4 eV. This stands in sharp contrast to the values of $U_c$, typically ~2-3 eV, found in pressurized La$_3$Ni$_2$O$_7$ phases exhibiting high-Tc superconductivity. This significant reduction in the critical $U_c$ signifies a substantially enhanced magnetic instability within the La$_2$NiO$_4$ structure relative to its bilayer counterpart. The La$_2$NiO$_4$ system exhibits a marked predisposition towards magnetic ordering, persisting down to ultralow interaction strengths.

We attribute this pronounced magnetic instability to two primary factors: Firstly, reduced dimensionality in the monolayer La$_2$NiO$_4$ structure likely leads to weaker electronic screening compared to the bilayer La$_3$Ni$_2$O$_7$ system. Consequently, the effective on-site Coulomb repulsion $U$ within La$_2$NiO$_4$ is presumably larger than that realized in La$_3$Ni$_2$O$_7$ under comparable conditions. Secondly, the ultralow $U_c$ value intrinsically favors magnetic instability. The convergence of these effects, the increased effective interaction strength combined with a drastically reduced instability threshold, strongly disfavors superconductivity. Instead, it robustly stabilizes the observed magnetic ordered state as the dominant ground state configuration in La$_2$NiO$_4$ [49-53].

**Potential Superconductivity in La$_2$NiO$_4$** To explore the potential for superconductivity of this system, we constructed the RPA linearized gap equation below the critical interaction strength $U_c$ to solve for the dominant superconducting pairing instability. The non-interacting spin susceptibility $\chi_0(q, \omega = 0)$, and the RPA-enhanced spin susceptibility $\chi_{RPA}(q, \omega = 0)$ were computed at T=50K, as displayed in Fig 4(a,b), respectively.

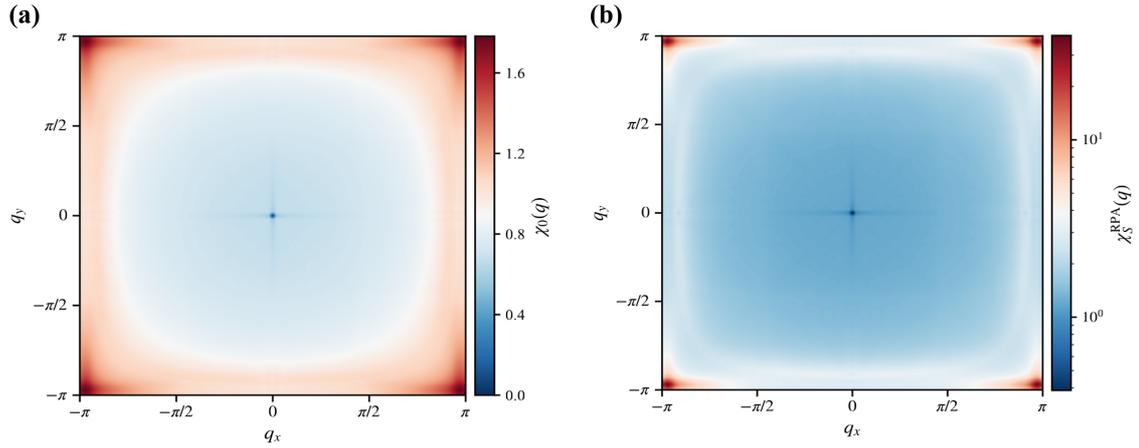

Fig 4. Non-interacting spin susceptibility $\chi_0(q, \omega = 0)$ (a) and correlation-enhanced RPA spin susceptibility $\chi_{RPA}(q, \omega = 0)$ (b) at ambient pressure, plotted in linear and logarithmic scales, respectively.

As clearly evident from Fig. 4(a,b), the RPA spin susceptibility exhibits a pronounced peak structure at the antiferromagnetic wavevector q=($\pi$,$\pi$). The magnitude of this peak significantly exceeds the susceptibility found in other regions of the Brillouin zone. This striking enhancement at ($\pi$,$\pi$) arises directly from the amplification effect of the Coulomb interaction $U$, which markedly enhances spin fluctuations specifically at this wavevector.

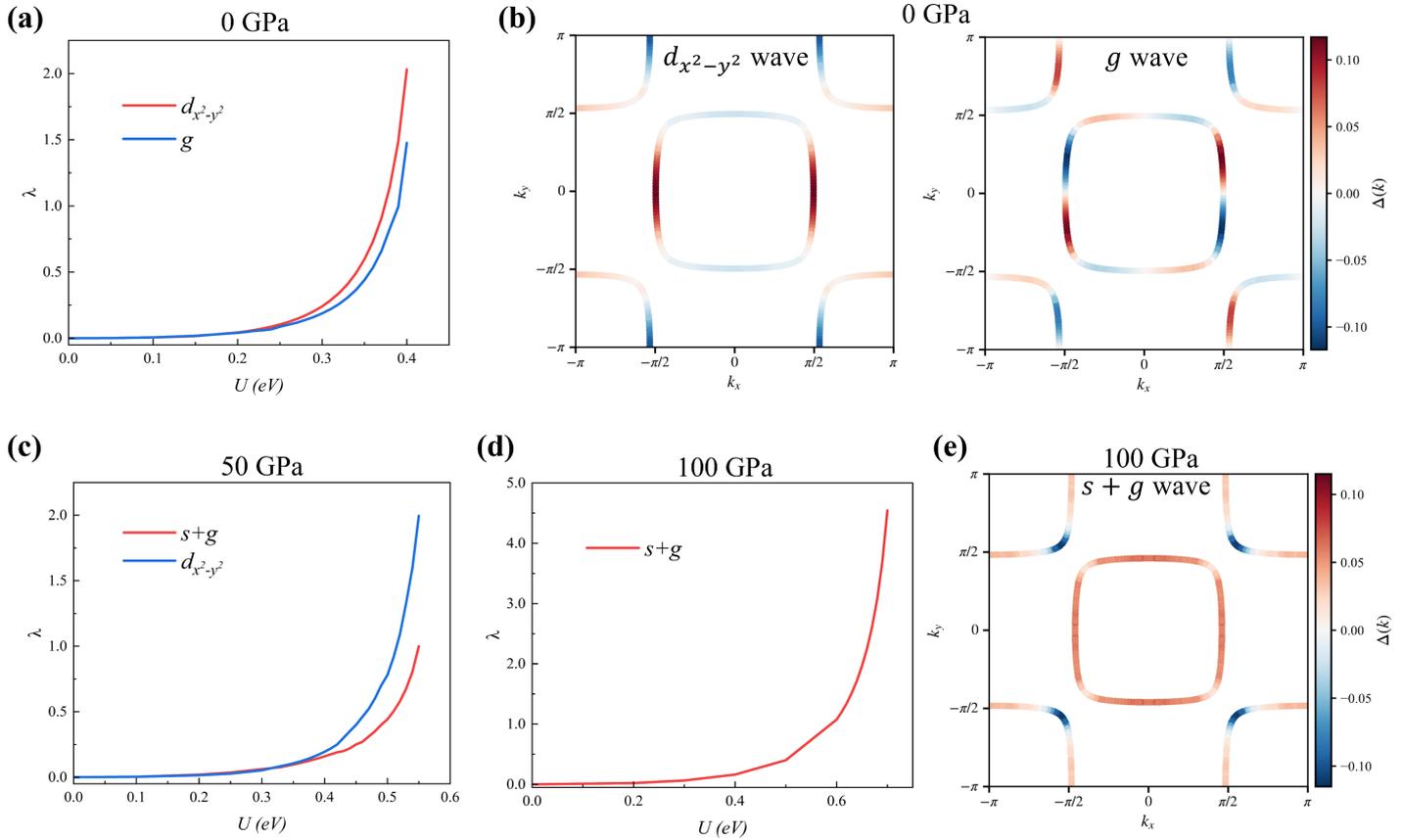

Fig. 5 Evolution of pairing interaction strength $\lambda$ with Coulomb interaction $U$ at ambient pressure(a) and momentum-space distribution of the dominant gap function(b), exhibiting $d_{x^2-y^2}$-wave and $g$-wave symmetries, respectively. Dependence of pairing interaction strength $\lambda$ on $U$ at 50 GPa(c). Evolution of pairing interaction strength $\lambda$ with $U$ at 100 GPa(d), and momentum-space distribution of the dominant gap function(e), exhibiting the $s+g$-wave symmetry, which manifests distinct symmetry characters across Fermi pockets.

At ambient pressure, within the parameter range of $U = 0 - 0.4\,eV$, $U' = U - 2J_H$, and $J_H = J_P = 0 - 0.1\,eV$, the superconducting gap function exhibits a predominantly $d_{x^2-y^2}$-wave symmetry, as depicted in Fig 5(a,b), respectively. Crucially, an increasing $J_H$ leads to an admixture of approximately 5% to 20% $g$-wave symmetry character in the gap structure. We identify a subleading superconducting instability with $g$-wave symmetry, exhibiting a pairing strength comparable to that of the dominant $d_{x^2-y^2}$-wave channel, while all other pairing symmetries exhibit significantly weaker superconducting instabilities. This progressive $g$-wave mixing signifies a notable modification in the pairing mechanism's sensitivity to anisotropy. Specifically, the robust dominance of $d_{x^2-y^2}$-wave pairing is universally understood as strong evidence for a superconducting instability driven primarily by spin fluctuations peaked at the antiferromagnetic wavevector q=(π,π)—the characteristic of the near-neighbor antiferromagnetic correlations. The emergence and amplification of the $g$-wave component with increasing Hund's coupling $J_H$, however, highlights the competing role of pair-breaking scattering channels. Enhanced $J_H$ generally strengthens inter-orbital spin-flip and pair-hopping processes, which tends to suppress anisotropic gap structures and favors more isotropic pairing. The induced $g$-wave admixture likely reflects a mild but distinct competition between the primary $d_{x^2-y^2}$ magnetic pairing channel and Hund's coupling-induced scattering near specific Fermi surface loci, potentially associated with orbital texture variations or nesting conditions favoring higher angular momentum contributions.

These results indicate that as pressure increases, the dominant $d$-wave and $g$-wave gap symmetries at ambient pressure are progressively suppressed. Concurrently, a novel superconducting gap with the $s+g$-wave mixed symmetry emerges and becomes stronger and stronger. Its associated pairing strength $\lambda$ gradually increases and eventually surpasses that of the $d_{x^2-y^2}$-wave and $g$-wave symmetries at approximately 75 GPa. By 100 GPa, the $s+g$-wave mixed symmetry completely dominates the superconducting state. The evolution of $\lambda$ at 50 GPa and 100 GPa as a function of the Hubbard $U$ is shown in Fig. 5(c, d), respectively. Upon elevating pressure to 100 GPa and employing interaction parameters $U = 0 - 0.6\,eV$, the superconducting gap symmetry undergoes a pronounced transformation. As revealed in Fig 5(d) and (e), the dominant pairing symmetry transitions to an $s+g$-wave component, whereas all other competing symmetries exhibit notably weaker pairing strengths. Notably, the gap symmetry manifests Fermi surface-dependent anisotropy: Near the $\Gamma$-point Fermi pocket, a nodeless $s$-wave gap dominates, whereas in the vicinity of the M(π,π)-point Fermi sheet, the gap displays pronounced $g$-wave anisotropy with sign-changing nodes.

**Impact of Self-Doping Effects** As shown in Fig. 2(a-c), within the DFT band structures at various pressures, we previously highlighted that when pressure exceeds 25 GPa, a band originating from La-f electrons crosses the Fermi level, forming a small electron pocket near the Z point. This self-doping effect leads to a deviation from the pristine half-filling condition in the electron occupation of the Ni-$3d$ $e_g$ bands near the Fermi level, which are crucial for superconducting pairing, thereby affecting the superconducting pairing potential. To quantitatively investigate the impact of self-doping on the pairing mechanism, we constructed a theoretical model at 100 GPa simulating the exact half-filling scenario i.e. electron occupation number = 1 without self-doping influence. The Hamiltonian of this model retains the form of Eq. 2, and the employed tight-binding parameters are listed in Table 3. Nearest-neighbor hopping terms are derived from Wannierization of the DFT band structures at 100 GPa. On-site energies are adjusted to enforce an electron occupancy of 1.

Table 3. Tight binding hopping parameters $t_{ij,ll'}$ for the theoretical model without self-doping under 100 GPa.

|  | $t_x^0$ | $t_x^1$ | $t_x^2$ | $t_x^3$ | $t_z^0$ | $t_z^1$ | $t_{xz}^{1,x}$ | $t_{xz}^{1,y}$ |
|---|---|---|---|---|---|---|---|---|
| $t_{ij,ll'}$ | 0.278 | -0.398 | 0.071 | -0.020 | -0.176 | -0.072 | -0.163 | 0.163 |

Performing the RPA calculations on this theoretical model, we first observe a critical Stoner instability interaction strength $U_c$ of 0.774, slightly larger than the value of 0.724 obtained for the realistic model including self-doping under 100 GPa. This indicates that the self-doping effect facilitates the emergence of magnetic order in La$_2$NiO$_4$—a feature that, in high-Tc systems, often implies a reduction in the phase space available for superconductivity, though the magnitude of this effect appears moderate in the present case. Furthermore, as illustrated in Fig. 6, below $U_c$, the dominant superconducting pairing symmetries revert to $d_{x^2-y^2}$-wave and $g$-wave—the same as those identified at ambient pressure—once the self-doping effect is removed. Aside from the difference in the required interaction strength $U$ for the divergence of the pairing strength $\lambda$, and the fact that the $\lambda$ of the $g$-wave channel becomes closer to that of the dominant $d_{x^2-y^2}$-wave, the structure of the $d_{x^2-y^2}$-wave and $g$-wave channels closely resemble those observed at ambient pressure, as seen in Fig. 4 (b). These results strongly suggest that the deviation in electron occupancy induced by self-doping is the primary cause for the suppression of $d$-wave dominance in the high-pressure regime. It is noteworthy that the restored $d_{x^2-y^2}$ dominance in the absence of self-doping aligns with the common symmetry framework in cuprate superconductors [54], underscoring a potential universal pairing mechanism rooted in strong electronic correlations within the NiO$_2$ planes. The fact that a relatively small change in occupancy—driven by the La-$f$ band crossing—can alter the leading pairing channel highlights the sensitive interplay between doping, Fermi surface topology, and correlation-driven instabilities in nickelates. Further studies exploring how such subtle occupancy shifts interact with other factors, such as interlayer coupling and orbital hybridization, could provide deeper insight into tailoring superconducting properties in these materials.

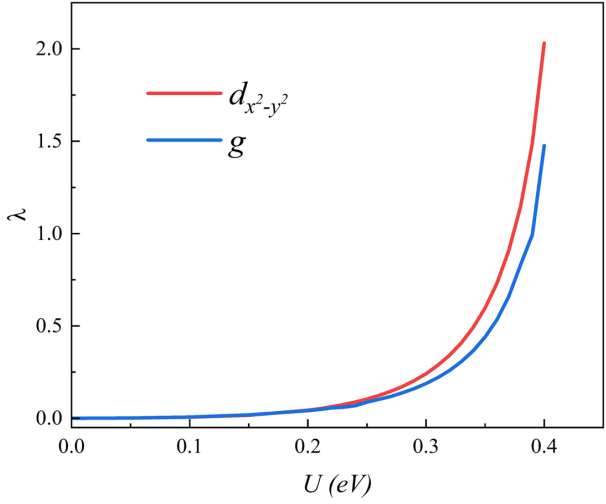

Fig. 6 Evolution of pairing interaction strength $\lambda$ with Coulomb interaction $U$ of the theoretical model in Tab. 3 without self-doping under 100 GPa

**Discussion and Conclusion** This study presents a comprehensive investigation on the electronic structures, magnetic instabilities, and superconducting property of monolayered nickelate $La_2NiO_4$ under hydrostatic pressure. The DFT+DMFT calculations reveal pronounced coherence dissolution near the Fermi level in $La_2NiO_4$ at low pressures, exhibiting distinct non-Fermi-liquid behavior. Crucially, multi-orbital susceptibility analysis reveals an exceptionally low critical Stoner parameter $U_c$, indicating robust magnetic order that competitively precludes superconductivity in the pristine system. Below this $U_c$ threshold, however, emergent superconducting instabilities exhibit pressure-tuned symmetry evolution: $d_{x^2-y^2}$-wave pairing dominates at ambient pressure, while $s + g$-wave symmetry prevails above 75 GPa, mediated by renormalized spin-fluctuation channels.

We also find that while monolayer $La_2NiO_4$ exhibits low-energy electronic structures near Fermi level remarkably similar to its bilayer counterpart $La_3Ni_2O_7$ [35-38], the absence of observable superconductivity in $La_2NiO_4$ stems from two interrelated factors: the extreme stability of its magnetic ground state, evidenced by a critical Stoner instability interaction strength $U_c$, $U_c/W \approx 0.15$ at 100 GPa versus 0.4–0.5 in $La_3Ni_2O_7$ at 15 GPa [44], necessitates prohibitively high pressures to achieve magnetic suppression within experimental limits; concurrently, even below Uc, hydrostatic pressure actively destabilizes the optimal $d_{x^2-y^2}$-wave pairing symmetry, forcing a transition into energetically unfavorable $s + g$-wave symmetry that rarely manifests in high-Tc systems due to intrinsic high-angular-momentum penalties and competition with alternative quantum orders. Consequently, realizing unconventional superconductivity in La$_2$NiO$_4$ demands ambient-pressure strategies, such as targeted chemical doping or epitaxial strain, to selectively suppress magnetic order while preserving strong spin fluctuations, circumventing the detrimental high-pressure pathway that compromises pairing symmetry viability.

**Acknowledgements:**

This work is supported by the National Natural Science Foundation of China (NSFC) under Grant Nos. 11974354, U2030114 and the CASHIPS Director's Fund under Grant No. YZJJ202207-CX. Numerical calculations were partly performed in the Center for Computational Science of CASHIPS, the ScGrid of Supercomputing Center, Computer Network Information Center of Chinese Academy of Sciences, and the Hefei Advanced Computing Center.